\documentclass[12pt]{article}
\usepackage{myart}

\normalbaselines
\input tcilatex
\begin{document}

\author{Yuri A. Rylov}
\title{On strategy of relativistic quantum theory construction}
\date{Institute for Problems in Mechanics, Russian Academy of Sciences,\\
101-1, Vernadskii Ave., Moscow, 119526, Russia.\\
e-mail: rylov@ipmnet.ru\\
Web site: {$http://rsfq1.physics.sunysb.edu/\symbol{126}rylov/yrylov.htm$}\\
or mirror Web site: {$http://gasdyn-ipm.ipmnet.ru/\symbol{126}%
rylov/yrylov.htm$}}
\maketitle

\begin{abstract}
Two different strategies of the relativistic quantum theory construction are
considered and evaluated. The first strategy is the conventional strategy,
based on application of the quantum mechanics technique to relativistic
systems. This approach cannot solve the problem of pair production. The
apparent success of QFT at solution of this problem is conditioned by the
inconsistency of QFT, when the commutation relations are incompatible with
the dynamic equations. (The inconsistent theory "can solve" practically any
problem, including the problem of pair production). The second strategy is
based on application of fundamental principles of classical dynamics and
those of statistical description to relativistic dynamic systems. It seems
to be more reliable, because this strategy does not use quantum principles,
and the main problem of QFT (join of nonrelativistic quantum principles with
the principles of relativity) appears to be eliminated.
\end{abstract}

\section{Introduction}

The conventional quantum theory has been tested very well only for
nonrelativistic physical phenomena of microcosm. The quantum theory is
founded on the nonrelativistic quantum principles. Application of the
quantum theory to relativistic phenomena of microcosm meets the problem of
join of the nonrelativistic quantum principles with the principles of the
relativity theory. Many researchers believe that such a join has been
carried out in the relativistic quantum field theory (QFT). Unfortunately,
it is not so, and the main difficulty lies in the fact that we do not apply
properly the relativity principles. Writing dynamic equations in the
relativistically covariant form, we believe that we have taken into account
all demands of the relativity theory.

Unfortunately, it is not so. The relativistic invariance of dynamic
equations is the necessary condition of true application of the relativistic
principles, but it is not sufficient. Besides, it is necessary to use the
relativistic concept of the state of the considered physical objects. For
instance, describing a particle in the nonrelativistic mechanics, we
consider the pointlike particle in the three-dimensional space to be a
physical object, whose state is described by the particle position $\mathbf{x%
}$ and momentum $\mathbf{p}$. The world line of the particle is considered
to be a history of the particle, but not a physical object. However, in the
relativistic mechanics the particle world line $\mathcal{L}$ is considered
to be a physical object (but not its history). In this case the pointlike
particle is an intersection $\mathcal{L}\cap \mathcal{T}_{C}$ of the world
line $\mathcal{L}$ with the hyperplane $\mathcal{T}_{C}$:$\ \ t=C=$const.
The hyperplane $\mathcal{T}_{C}$ is not invariant in the sense, that the set 
$\mathcal{S}_{T}$ of all hyperplanes $\mathcal{T}_{C}$ is not invariant with
respect to the Lorentz transformations. If we have several world lines $%
\mathcal{L}_{1}$, $\mathcal{L}_{2}$, ... $\mathcal{L}_{n}$, their
intersections $\mathcal{L}_{k}\cap \mathcal{T}_{C}$ with the hyperplane $%
\mathcal{T}_{C}$ form a set $\mathcal{S}_{C}$ of particles $\mathcal{P}_{k}=%
\mathcal{L}_{k}\cap \mathcal{T}_{C}$ in some coordinate system $K$. The set $%
\mathcal{S}_{C}$ of particles $\mathcal{P}_{k}$ depends on the choice of the
coordinate system $K$. In the coordinate system $K^{\prime }$, moving with
respect to the coordinate system $K,$ we obtain another set $\mathcal{S}%
_{C^{\prime }}^{\prime }$ of particles $\mathcal{P}_{k}^{\prime }=\mathcal{L}%
_{k}\cap \mathcal{T}_{C^{\prime }}^{\prime }$, taken at some time moment $%
t^{\prime }=C^{\prime }=$const. If we have only one world line $\mathcal{L}%
_{1}$, we may choose the constant $C^{\prime }$ in such a way, that $%
\mathcal{P}_{1}=\mathcal{L}_{1}\cap \mathcal{T}_{C}$ coincides with $%
\mathcal{P}_{1}^{\prime }=\mathcal{L}_{1}\cap \mathcal{T}_{C^{\prime
}}^{\prime }$. However, if we have many world lines coincidence of sets $%
\mathcal{S}_{C}$ and $\mathcal{S}_{C^{\prime }}^{\prime }$ is impossible at
any choice of the constant $C^{\prime }$. In other words, the particle is
not an invariant object, and it cannot be considered as a physical object in
the relativistic mechanics. In the case of one world line we can compensate
the noninvariant character of a particle by a proper choice of the constant $%
C^{\prime }$, but at the statistical description, where we deal with many
world lines, such a compensation is impossible.

In the nonrelativistic mechanics there is the absolute simultaneity, and the
set $\mathcal{S}_{C}$ hyperplanes $\mathcal{T}_{C}:t=$const is the same in
all inertial coordinate systems. In this case intersections of world lines $%
\mathcal{L}_{1}$, $\mathcal{L}_{2}$, ... $\mathcal{L}_{n}$ with the
hyperplane $\mathcal{T}_{C}$ form the same set of events in all coordinate
systems, and particles are invariant objects, which may be considered as
physical objects. Strictly, the world line should be considered as a
physical object also in the nonrelativistic physics, as far as the
nonrelativistic physics is a special case of the relativistic one. But, in
this case a consideration of a particle as a physical object is also
possible, and this consideration is simpler and more effective, as far as
the pointlike particle is a simpler object, than the world line.

The above statements are not new. For instance, V.A. Fock stressed in his
book \cite{F55}, that concept of the particle state is different in
relativistic and nonrelativistic mechanics. As a rule researchers do not
object to such statements, but they do not apply them in practice. The
nonrelativistic quantum theory has been constructed and tested in many
experiments. It is a starting point for construction of the relativistic
quantum theory. In this paper we try to investigate different strategies of
the relativistic quantum theory construction, in order to choose the true
one. However, at first we consider interplay between the fundamental
physical theory and the truncated physical theory.

The scheme of this interplay is shown in the figure. The fundamental theory
is a logical structure. The fundamental principles of the theory are shown
below. The experimental data, which are to be explained by the theory are
placed on high. Between them there is a set of logical corollaries of the
fundamental principles. It is possible such a situation, when for some
conditions one can obtain a list of logical corollaries, placed near the
experimental data. It is possible such a situation, when some circle of
experimental data and of physical phenomena may be explained and calculated
on the basis of this list of corollaries without a reference to the
fundamental principles. In this case the list of corollaries of the
fundamental principles may be considered as an independent physical theory.
Such a theory will be referred to as the truncated theory, because it
explains not all phenomena, but only a restricted circle of these phenomena
(for instance, only nonrelativistic phenomena). Examples of truncated
physical theories are known in the history of physics. For instance, the
thermodynamics is such a truncated theory, which is valid only for the
quasi-static thermal phenomena. The thermodynamics is an axiomatic theory.
It cannot be applied to nonstationary thermal phenomena. In this case one
should use the kinetic theory, which is a more fundamental theory, as far as
it may be applied to both quasi-static and nonstationary thermal phenomena.
Besides, under some conditions the thermodynamics can be derived from the
kinetic theory as a partial case.

The truncated theory has a set of properties, which provide its wide
application.

\begin{enumerate}
\item The truncated theory is simpler, than the fundamental one, because a
part of logical reasonings and mathematical calculations of the fundamental
theory are used in the truncated theory in the prepared form. Besides, the
truncated theory is located near experimental data, and one does not need
long logical reasonings for application of the truncated theory.

\item The truncated theory is a list of prescriptions, and it is not a
logical structure in such extent, as the fundamental theory is a logical
structure. The truncated theory is axiomatic, it contains more axioms, than
the fundamental theory, as far as logical corollaries of the fundamental
theory appear in the truncated theory as fundamental principles (axioms).

\item Being simpler, the truncated theory appears before the fundamental
theory. It is a reason of conflicts between the advocates of the fundamental
theory and advocates of the truncated theory, because the last consider the
truncated theory to be the fundamental one. Such a situation took place, for
instance, at becoming of the statistical physics, when advocates of the
axiomatic thermodynamics oppugn against Gibbs and Boltzmann. Such a
situation took place at becoming of the doctrine of
Copernicus-Galileo-Newton, when advocates of the Ptolemaic doctrine oppugn
against the doctrine of Copernicus-Galileo-Newton. They referred that there
was no necessity to introduce the Copernican doctrine, as far as the
Ptolemaic doctrine is simple and customary. Only discovery of the Newtonian
gravitation law and consideration of the celestial phenomena, which cannot
be described in the framework of the Ptolemaic doctrine, terminated the
contest of the two doctrines.

\item Constructing the truncated theory before the fundamental one, the
trial and error method is used usually. In other words, the truncated theory
is guessed, but not constructed by a logical way.
\end{enumerate}

The main defect of the truncated theory is an impossibility of its expansion
over wider circle of physical phenomena. For instance, let the truncated
theory explain nonrelativistic physical phenomena. It means, that the basic
propositions of the truncated theory are obtained as corollaries of the
fundamental principles and of nonrelativistic character of the considered
phenomena. To expand the truncated theory on relativistic phenomena, one
needs to separate, what in the principles of the truncated theory is a
corollary of fundamental principles and what is a corollary of
nonrelativistic character of the considered phenomena. A successful
separation of the two factors means essentially a perception of the theory
truncation and construction of the fundamental theory. If the fundamental
theory has been constructed, the expansion of the theory on the relativistic
phenomena is obtained by an application of the fundamental principles to the
relativistic phenomena. The obtained theory will describe the relativistic
phenomena correctly. It may be distinguished essentially from the truncated
theory, which is applicable for description of only nonrelativistic
phenomena.

The conventional nonrelativistic quantum theory is a truncated theory, which
is applicable only for description of nonrelativistic phenomena. It has
formal signs of the truncated theory (long list of axioms, simplicity,
nearness to experimental data). Truncated character of the nonrelativistic
quantum theory is called in question usually by researchers working in the
field of the quantum theory. The principal problem of the relativistic
quantum theory is formulated usually as a problem of unification of the
nonrelativistic quantum principles with the principles of the relativity
theory.

\textit{Conventionally the nonrelativistic quantum theory is considered to
be a fundamental theory}. The relativistic quantum theory is tried to be
constructed without puzzling out, what in the nonrelativistic quantum theory
is conditioned by the fundamental principles and what is conditioned by its
nonrelativistic character. It is suggested that the linearity is the
principal property of the quantum theory, and it is tried to be saved.
However, the analysis shows that the linearity of the quantum theory is some
artificial circumstance \cite{R2005}, which simplifies essentially the
description of quantum phenomena, but it does not express the essence of
these phenomena. The conventional approach to construction of the
relativistic quantum theory is shown by the dashed line in the scheme.
Following this line, the construction of the true relativistic quantum
theory appears to be as difficult, as a discovery of the Newtonian
gravitation law on the basis of the Ptolemaic conception, because in this
case only the trial and errors method can be used. Besides, even if we
succeeded to construct such a theory, it will be very difficult to choose
the valid version of the theory, because it has no logical foundation. In
other words, the conventional approach to construction of the relativistic
quantum theory (invention of new hypotheses and fitting) seems to lead to
blind alley, although one cannot eliminate the case that it appears to be
successful. (the trial and error method appeared to be successful at
construction of the nonrelativistic quantum mechanics).

Alternative way of construction of the relativistic theory of physical
phenomena in the microcosm is shown in figure by the solid line. It supposes
a derivation of fundamental principles and their subsequent application to
the relativistic physical phenomena. Elimination of the nonrelativistic
quantum principles is characteristic for this approach. This elimination is
accompanied by elimination of the problem of an unification of the
nonrelativistic quantum principles with the relativity principles.
Simultaneously one develops dynamic methods of the quantum system
investigation, when the quantum system is investigated simply as a dynamic
system. These methods are free of application of quantum principles. They
are used for investigation of both relativistic and nonrelativistic quantum
systems. A use of logical constructions is characteristic for this approach.
One does not use an invention of new hypotheses and fitting (the trial and
error method).

It is assumed usually that quantum systems contain such a specific
nonclassical object as the wave function. Quantum principles is a list of
prescriptions, how to work with the wave functions. In reality the wave
function is not a specific nonclassical object. The wave function is a
complex hydrodynamic potential. Any ideal fluid can be described in terms
the hydrodynamic potentials (Clebsch potentials \cite{C57,C59}). In
particular, it can be described in terms of the wave functions \cite{R99}.
Prescriptions for work with description in terms of wave functions follows
directly from definition of the wave function and from prescriptions for
work with the dynamic systems of hydrodynamic type. Quantum systems are such
dynamic systems of hydrodynamic type, for which the dynamic equations are
linear, if they are written in terms of the wave function. Statistical
ensemble $\mathcal{E}\left[ \mathcal{S}_{\mathrm{st}}\right] $ of stochastic
particles $\mathcal{S}_{\mathrm{st}}$ is a dynamic system of hydrodynamic
type. Under some conditions dynamic equations for the statistical ensemble $%
\mathcal{E}\left[ \mathcal{S}_{\mathrm{st}}\right] $ become linear, if they
are written in terms of the wave function. In this case the statistical
ensemble $\mathcal{E}\left[ \mathcal{S}_{\mathrm{st}}\right] $ may be
considered as a quantum system in the sense, that quantum principles (the
prescriptions for work with wave function) may be applied to the statistical
ensemble $\mathcal{E}\left[ \mathcal{S}_{\mathrm{st}}\right] $.

Thus, the quantum systems are not enigmatic systems, described by a specific
nonclassical object (wave function). Quantum systems are a partial case of
dynamic systems, which may and must be investigated by conventional dynamic
methods, applied in the fluid dynamics. The classical principles of dynamics
and those of statistical description are fundamental principles of any
dynamics and, in particular, of the quantum mechanics, considered to be a
dynamics of stochastic particles. In other words, the nonrelativistic
quantum theory is truncated theory with respect to dynamics of the
stochastic systems.

Transition to relativistic quantum mechanics means that one should apply the
general principles of mechanics to the statistical ensembles of stochastic
particles, whose regular component of velocity is relativistic. (Stochastic
component of velocity is always relativistic, even in the case, when the
regular component is nonrelativistic). Such a statistical description can be
carried out in terms of the wave function. However, we cannot state
previously that dynamic equations will be linear, because in the
relativistic case there is such a phenomenon as the particle production,
which is absent in the classical relativistic mechanics and in the
nonrelativistic quantum theory.

At first sight, the direct way of transition from nonrelativistic quantum
theory to the relativistic one seems to be more attractive, because it is
simpler and it does not need a discovery of fundamental concepts. Besides,
it seems to be an unique way, if we believe that the nonrelativistic quantum
theory is a fundamental theory (but not a truncated one). Unfortunately,
following the quantum principles and this way, we come to a blind alley.
This circumstance forces us to question, whether the nonrelativistic quantum
theory is a fundamental theory (but not a truncated one).

We shall refer to the path, shown by the dashed line as the direct path
(direct approach). The path, shown by the solid line will be referred to as
the logical path (logical approach). Investigation of the two approaches and
of investigation strategies connected with them is the main goal of this
paper. The logical path seems to be more adequate, but at the same time it
seems to be more difficult. There are two different methods of presentations
of our investigation: (1) description of problems of the direct path from
the viewpoint of the logical one, (2) description of those problems of the
direct path which have lead to a refusal from the direct path in favour of
the logical one. In this paper we prefer to use the second version.

\section{Difficulties of the quantum principles \newline
application to the relativistic phenomena}

The particle production is the physical phenomenon, which is characteristic
only for quantum relativistic physics. This phenomenon has no classical
analog, because it is absent in the classical relativistic physics. This
phenomenon is absent in the nonrelativistic quantum physics. At the
classical description the particle production is a turn of the world line in
the time direction. According to such a conception the particles are
produced by pairs particle -- antiparticle. In classical physics there is no
force field, which could produce or annihilate such pairs. If the world line
describes the pair production, some segment of this world line is to be
spacelike. At this segment we have

\begin{equation}
g_{ik}\frac{dx^{i}}{d\tau }\frac{dx^{k}}{d\tau }<0  \label{b1.1}
\end{equation}%
where $g_{ik}$ is the metric tensor and $x^{k}=x^{k}\left( \tau \right) $ is
the equation of the world line $\mathcal{L}$. On the other hand, the action
for the free classical relativistic particle has the form%
\begin{equation}
\mathcal{A}\left[ x\right] =-\int mc\sqrt{g_{ik}\frac{dx^{i}}{d\tau }\frac{%
dx^{k}}{d\tau }}d\tau  \label{b1.2}
\end{equation}%
Relations (\ref{b1.1}) and (\ref{b1.2}) are incompatible. They become
compatible, if there is such a force field, which changes the particle mass.
For instance, if instead of the action (\ref{b1.2}) we have 
\begin{equation}
\mathcal{A}\left[ x\right] =\int L\left( x,\dot{x}\right) d\tau ,\qquad
L=-m_{\mathrm{eff}}c\sqrt{g_{ik}\frac{dx^{i}}{d\tau }\frac{dx^{k}}{d\tau }}%
,\qquad m_{\mathrm{eff}}=m\sqrt{1+f}  \label{b1.3}
\end{equation}%
where $m_{\mathrm{eff}}$ is the effective particle mass, and $f=f\left(
x\right) $ is some external force field, which changes the effective
particle mass $m_{\mathrm{eff}}$. If $f<-1$, the effective mass is
imaginary, the condition (\ref{b1.1}) takes place in the region, where $f<-1$%
, and the interdict on the pair production, or on the pair annihilation is
violated.

Further we shall use the special term WL for the world line considered as a
physical object. The term "WL" is the abbreviation of the term "world line".
Along with the term WL we shall use the term "emlon", which is the perusal
of Russian abbreviation "ML", which means world line. Investigation of the
emlon, changing its direction in the time direction and describing pair
production (or annihilation), shows \cite{R70,R2003}, that some segments of
the emlon describe a particle, whereas another segments describe an
antiparticle. The particle and antiparticle have opposite sign of the
electric charge. The energy $E=\int T^{00}d\mathbf{x}$ of the particle and
that of the antiparticle is always positive. The time components $%
p_{0}=\partial L/\partial \dot{x}^{0}$, $\dot{x}^{k}\equiv dx^{k}/d\tau $ of
the canonical momentum $p_{k}$ of the particle and that of the antiparticle
have opposite sign, if the world line (WL) is considered as a single
physical object (single dynamic system). They may have the same sign and
coincide with $-E$, if different segments of the emlon are associated with
different dynamic systems (particles and antiparticles).

Description of the annihilation process as an evolution of two different
dynamic systems (particle and antiparticle), which cease to exist after
collision, is incompatible with the conventional formalism of classical
relativistic dynamics, where dynamic systems may not disappear. However,
description of this process as an evolution of some pointlike object SWL
moving along the world line in the direction of increase of the evolution
parameter $\tau $ is possible from viewpoint of the conventional formalism
of the relativistic physics. The object SWL is the abbreviation of the term
"section of world line". Along with the term "SWL" we shall use also the
term "esemlon". It is the perusal of Russian abbreviation "SML", which means
"Section of the world line". The esemlon is the collective concept with
respect to concepts of particle and antiparticle. In the process of
evolution the esemlon may change its state (particle or antiparticle). Such
an approach is compatible with the relativistic kinematics.

The investigation \cite{R70} shows that the energy $E$ and the temporal
component of the canonical momentum $-p_{0}$ are different quantities, which
may coincide, only if there is no pair production. In the presence of pair
production the equality $E=-p_{0}$ for the whole world line is possible also
in the case, when the whole world line is cut into segments, corresponding
to particles and antiparticles, and each segment is considered to be a
single dynamic system.

It is generally assumed that the perturbation theory and the divergencies
are the main problems of QFT. In reality, it is only a vertex of iceberg.
The main problem lies in the definition of the commutation relations. We
demonstrate this in the example of the dynamic equation%
\begin{equation}
(\partial _{i}\partial ^{i}+m^{2})\varphi =\lambda \varphi ^{\ast }\varphi
\varphi  \label{b3.1}
\end{equation}%
Here $\varphi $ is the scalar complex field and $\varphi ^{\ast }$ is the
Hermitian conjugate field, $\lambda $ is the self-action constant. There are
two different schemes of the second quantization: (1) $PA$-scheme, where
particle and antiparticle are considered as different physical objects and
(2) $WL$-scheme, where world line (WL) is considered as a physical object. A
particle and an antiparticle are two different states of SWL (or WL). These
two schemes distinguish in the commutation relations, imposed on the
operators $\varphi $ and $\varphi ^{\ast }$ (see for details \cite{R2001}).

In the $PA$-scheme there is indefinite number of objects (particles and
antiparticles) which can be produced and annihilated. The commutators $\left[
\varphi \left( x\right) ,\varphi \left( x^{\prime }\right) \right] _{-}$ and 
$\left[ \varphi \left( x\right) ,\varphi ^{\ast }\left( x^{\prime }\right) %
\right] _{-}$ vanish%
\begin{equation}
\left[ \varphi \left( x\right) ,\varphi \left( x^{\prime }\right) \right]
_{-}=0,\qquad \left[ \varphi \left( x\right) ,\varphi ^{\ast }\left(
x^{\prime }\right) \right] =0,\qquad \left\vert x-x^{\prime }\right\vert
^{2}<0  \label{b3.2}
\end{equation}%
if interval between the points $x$ and $x^{\prime }$ is spacelike. The $PA$%
-scheme tried to describe the entire picture of the particle motion and
their collision. It is a very complicated picture. It can be described only
in terms of the perturbation theory, because the number of physical objects
(objects of quantization) is not conserved. The commutation relation, which
are used in the $PA$-scheme are \textit{incompatible with the dynamic
equations}. As a result the $PA$-scheme of the second quantization appears
to be inconsistent.

In the $WL$-scheme of the second quantization the number of objects of
quantization (WL) is conserved, and one can divide the whole problem into
parts, containing one WL, two WLs, three WLs etc. Each of parts may be
considered and solved independently. The statement of the problem reminds
that of the nonrelativistic quantum mechanics, where the number of particles
is conserved. As a result the whole problem may be divided into one-particle
problem, two-particle problem, etc, and each problem can be solved
independently. According to such a division of the whole problem into
several simpler problems, the problem of the second quantization in $WL$%
-scheme is reduced to several simpler problems. As a result it may be
formulated without a use of the perturbation theory (see for details \cite%
{R2001}). Commutation relations in the $WL$-scheme do not satisfy the
condition (\ref{b3.2}). This circumstance is connected with the fact that
the objects of quantization (WLs) are lengthy objects. If there is the
particle production, WLs are spacelike in the sense that they may contain
points $x$ and $x^{\prime }$, separated by the spacelike interval. There are
such dynamic variables at $x$ and at $x^{\prime }$, lying on the same WL,
for which the commutator does not vanish, and it is a reason for violation
of conditions (\ref{b3.2}) in the $WL$-scheme of quantization. The
commutation relations in $WL$-scheme are compatible with dynamic equations.
Besides, simultaneous commutation relations depend on the self-action
constant $\lambda $. The $WL$-scheme of the second quantization is
consistent and compatible with dynamic equations. It can be solved by means
of nonperturbative methods. However, the pair production is absent in the $%
WL $-scheme, even if the self-action constant $\lambda \neq 0$ \cite{R2001}.

One believes, that there is the pair production in the $PA$-scheme. However,
the $PA$-scheme is inconsistent, and the pair production is a corollary of
this inconsistency \cite{R2001}. Thus, neither $PA$-scheme nor $WL$-scheme
of quantization can derive the pair production effect. It is connected with
the fact, that the self-action of the form (\ref{b3.1}), as well as other
power interactions cannot generate pair production. To generate the pair
production, one needs interaction of special type \cite{R2003}.

Advocates of the $PA$-scheme state that in the $WL$-scheme the causality
principle is violated, because the conditions (\ref{b3.2}) are not
fulfilled. It is not so, because the causality principle has the form (\ref%
{b3.2}) only for pointlike objects. For lengthy objects (spacelike world
lines) the causality principle has another form \cite{R2001}. The condition (%
\ref{b3.2}) states simply that the dynamic variables of different dynamic
systems commutate. But in the case of spacelike WL the points $x$ and $%
x^{\prime }$, separated by the spacelike interval, may belong to the same
dynamic system. In this case the condition (\ref{b3.2}) has to be violated.
But independently of whether or not the advocates of the $PA$-scheme are
right, the dynamic equation (\ref{b3.1}) does not describe the pair
production, and appearance of the pair production \cite{GJ68,GJ70,GJ970,GJ72}
is a result of incompatibility of the commutation relations with the dynamic
equation.

Conventionally one considers the commutation relations as a kind of initial
conditions for the dynamic equations. As a result one does not see a
necessity to test the compatibility of the commutation relations with the
dynamic equations. In reality the analogy between the commutation relations
and initial conditions is not true. The commutation relations are additional
constraints imposed on the dynamic variables. \textit{Compatibility of
additional constraints with the dynamic equations is to be tested.}
Dependence of the simultaneous commutation relations on the self-action
constant $\lambda $ in the $WL$-scheme, where such a compatibility takes
place, confirms the necessity of such a test.

Thus, a direct application of the quantum mechanics formalism to a
relativistic dynamic systems leads to impossibility of the particle
production description. It means that we should understand the essence of
the quantum mechanics formalism and revise its form in accordance with the
revised understanding of the quantum mechanics.

\section{Linearity of quantum mechanics}

To show that the linearity of quantum mechanics formalism is not an
essential inherent property of the fundamental theory, we consider the Schr%
\"{o}dinger particle, which is the dynamic system $\mathcal{S}_{\mathrm{S}}$
described by the action 
\begin{equation}
\mathcal{S}_{\mathrm{S}}:\qquad \mathcal{A}_{\mathrm{S}}\left[ \psi \right]
=\int \left\{ \frac{i\hbar }{2}\left( \psi ^{\ast }\partial _{0}\psi
-\partial _{0}\psi ^{\ast }\cdot \psi \right) -\frac{\hbar ^{2}}{2m}\mathbf{%
\nabla }\psi ^{\ast }\mathbf{\nabla }\psi \right\} d^{4}x  \label{c2.1}
\end{equation}%
where $\psi =\psi \left( t,\mathbf{x}\right) $ is a complex wave function.
The meaning of the wave function (connection between the particle and the
wave function) is described by the relations. 
\begin{equation}
\left\langle F\left( \mathbf{x},\mathbf{p}\right) \right\rangle =B\int \func{%
Re}\left\{ \psi ^{\ast }F\left( \mathbf{x},\mathbf{\hat{p}}\right) \psi
\right\} d\mathbf{x,\qquad \hat{p}}=-i\hbar \mathbf{\mathbf{\nabla },\qquad }%
B=\left( \int \psi ^{\ast }\psi d\mathbf{x}\right) ^{-1}  \label{c2.4}
\end{equation}%
which define the mean value $\left\langle F\left( \mathbf{x},\mathbf{p}%
\right) \right\rangle $ of any function $F\left( \mathbf{x},\mathbf{p}%
\right) $ of position $\mathbf{x}$ and momentum $\mathbf{p}$. We shall refer
to the relation (\ref{c2.4}) together with the restrictions imposed on its
applications as the quantum principles, because von Neumann \cite{N32} has
shown, that all propositions of quantum mechanics can be deduced from
relations of this type. Thus, the action (\ref{c2.1}) describes the quantum
mechanics formalism (dynamics), whereas the relation (\ref{c2.4}) forms a
basis for the conventional interpretation of the quantum mechanics.

Dynamic equation%
\begin{equation}
i\hbar \partial _{0}\psi =-\frac{\hbar ^{2}}{2m}\mathbf{\nabla }^{2}\psi ,
\label{c2.2}
\end{equation}%
generated by the action (\ref{c2.1}) is linear, and one believes that this
linearity together with the linear operators, describing all observable
quantities, is the inherent property of the quantum mechanics.

The quantum constant $\hbar $ is supposed to describe the quantum properties
in the sense, that setting $\hbar =0$ in the quantum description, we are to
obtain the classical description. However, setting $\hbar =0$ in the action (%
\ref{c2.1}), we obtain no description. All terms in the action contain $%
\hbar $, and it seems that the description by means of the action (\ref{c2.1}%
) is quantum from the beginning to the end. In reality the principal part of
the dynamic system $\mathcal{S}_{S}$ is classical, and the quantum
description forms only a part of the general description. In other words,
description in terms of the action (\ref{c2.1}) is an artificial description.

To show this, we transform the wave function $\psi $, changing its phase%
\begin{equation}
\psi \rightarrow \Psi _{b}:\quad \psi =\left\vert \Psi _{b}\right\vert \exp
\left( \frac{b}{\hbar }\log \frac{\Psi _{b}}{\left\vert \Psi _{b}\right\vert 
}\right) \quad b=\text{const}\neq 0  \label{c3.1}
\end{equation}%
Substituting (\ref{c3.1}) in the action (\ref{c2.1}), we obtain 
\[
\mathcal{A}_{\mathrm{S}}\left[ \Psi _{b}\right] =\int \left\{ \frac{ib}{2}%
\left( \Psi _{b}^{\ast }\partial _{0}\Psi _{b}-\partial _{0}\Psi _{b}^{\ast
}\cdot \Psi _{b}\right) -\frac{b^{2}}{2m}\mathbf{\nabla }\Psi _{b}^{\ast }%
\mathbf{\nabla }\Psi _{b}\right. 
\]%
\begin{equation}
+\left. \frac{b^{2}}{2m}\left( \mathbf{\nabla }\left\vert \Psi
_{b}\right\vert \right) ^{2}-\frac{\hbar ^{2}}{2m}\left( \mathbf{\nabla }%
\left\vert \Psi _{b}\right\vert \right) ^{2}\right\} dtd\mathbf{x}
\label{c3.2}
\end{equation}%
This change of variables leads to the replacement $\hbar \rightarrow b$ and
to appearance of two nonlinear terms which compensate each other, if $%
b=\hbar $. The change of variable does not change the dynamic system $%
\mathcal{S}_{\mathrm{S}}$, although the dynamic equation becomes nonlinear,
if $b^{2}\neq \hbar ^{2}$ 
\begin{equation}
ib\partial _{0}\Psi _{b}=-\frac{b^{2}}{2m}\mathbf{\nabla }^{2}\Psi _{b}-%
\frac{\hbar ^{2}-b^{2}}{8m}\left( \frac{\left( \mathbf{\nabla }\rho \right)
^{2}}{\rho ^{2}}+2\mathbf{\nabla }\frac{\mathbf{\nabla }\rho }{\rho }\right)
\Psi _{b},  \label{c3.4}
\end{equation}

However, the description in terms of $\Psi _{b}$ appears to be natural in
the sense, that after setting $\hbar =0$, the action $\mathcal{A}_{\mathrm{S}%
}\left[ \Psi _{b}\right] $ turns into the action 
\begin{equation}
\mathcal{A}_{\mathrm{Scl}}\left[ \Psi _{b}\right] =\int \left\{ \frac{ib}{2}%
\left( \Psi _{b}^{\ast }\partial _{0}\Psi _{b}-\partial _{0}\Psi _{b}^{\ast
}\cdot \Psi _{b}\right) -\frac{b^{2}}{2m}\mathbf{\nabla }\Psi _{b}^{\ast }%
\mathbf{\nabla }\Psi +\frac{b^{2}}{2m}\left( \mathbf{\nabla }\left\vert \Psi
_{b}\right\vert \right) ^{2}\right\} dtd\mathbf{x}  \label{c3.6a}
\end{equation}%
which carries out the classical description. It describes the statistical
ensemble $\mathcal{E}\left[ \mathcal{S}_{\mathrm{d}}\right] $ of free
classical particles $\mathcal{S}_{\mathrm{d}}$. The action $\mathcal{A}_{%
\mathcal{E}\left[ \mathcal{S}_{\mathrm{d}}\right] }$ for this statistical
ensemble can be represented in the form 
\begin{equation}
\mathcal{A}_{\mathcal{E}\left[ \mathcal{S}_{\mathrm{d}}\right] }\left[ 
\mathbf{x}\right] =\int \frac{m}{2}\left( \frac{d\mathbf{x}}{dt}\right)
^{2}dtd\mathbf{\xi }  \label{c3.6}
\end{equation}%
where $\mathbf{x}=\mathbf{x}\left( t,\mathbf{\xi }\right) $ is a 3-vector
function of independent variables $t,\mathbf{\xi =}\left\{ \xi _{1,}\xi
_{2},\xi _{3}\right\} $. The variables (Lagrangian coordinates) $\mathbf{\xi 
}$ label particles $\mathcal{S}_{\mathrm{d}}$ of the statistical ensemble $%
\mathcal{E}\left[ \mathcal{S}_{\mathrm{d}}\right] $. The statistical
ensemble $\mathcal{E}\left[ \mathcal{S}_{\mathrm{d}}\right] $ is a dynamic
system of the hydrodynamic type. One can show that the dynamic system,
described by the action $\mathcal{A}_{\mathrm{Scl}}\left[ \Psi _{b}\right] $
(\ref{c3.6a}) is a partial case (irrotational flow) of the dynamic system $%
\mathcal{E}\left[ \mathcal{S}_{\mathrm{d}}\right] $ \cite{R99}.

Connection between the Schr\"{o}dinger equation and hydrodynamic description
is well known \cite{M26,B52}. But a connection between the description in
terms of wave function and the hydrodynamic description was one-way. One can
transit from the Schr\"{o}dinger equation to the hydrodynamic equations, but
one cannot transit from hydrodynamic equations to the description in terms
of the wave function, because one \textit{needs} to \textit{integrate}
hydrodynamic equations. Indeed, the Schr\"{o}dinger equation consists of two
real first order equations for the density $\rho $ and the phase $\varphi $,
whereas the system of the hydrodynamic equations consists of four first
order equations for the density $\rho $ and for the velocity $\mathbf{v}$.
To obtain four hydrodynamic equations one needs to take gradient of the
equation for the phase $\varphi $ and introduce the velocity $\mathbf{v}%
=m^{-1}\mathbf{\nabla }\varphi $. On the contrary, if we transit from the
hydrodynamic description to the description in terms of the wave function,
we are to integrate hydrodynamic equations. In the general case this
integration was not known for a long time.

Change of variables, leading from the action $\mathcal{A}_{\mathcal{E}\left[ 
\mathcal{S}_{\mathrm{d}}\right] }\left[ \mathbf{x}\right] $ to the action $%
\mathcal{A}_{\mathrm{Scl}}\left[ \Psi _{b}\right] $ contains integration
(see \cite{R99} or mathematical appendices to papers \cite{R2004a,R2004b}).
The constant $b$ in the action $\mathcal{A}_{\mathrm{Scl}}\left[ \Psi _{b}%
\right] $ is an arbitrary constant of integration (gauge constant).
Arbitrary integration functions are "hidden" inside the wave function $\Psi
_{b}$. Thus, the limit of Schr\"{o}dinger particle (\ref{c3.2}) at $\hbar
\rightarrow 0$ is a statistical ensemble $\mathcal{E}\left[ \mathcal{S}_{%
\mathrm{d}}\right] $, but not an individual particle $\mathcal{S}_{\mathrm{d}%
}$. It means, that the \textit{wave\ function\ describes\ a statistical\
ensemble of\ particles,\ but\ not\ an\ individual particle}, and the
Copenhagen\ interpretation, where the wave function describes an individual
particle, is\ \textit{incompatible}\ with the\ quantum\ mechanics\ formalism.

Dynamic system $\mathcal{S}_{\mathrm{S}}$ is described by the action (\ref%
{c3.2}) as well as by the action (\ref{c2.1}). Interpretation (\ref{c2.4})
of the wave function $\psi $ may be also rewritten as interpretation of $%
\Psi _{b}$ by means of transformation (\ref{c3.1}). In the action (\ref{c3.2}%
) only one term contains the quantum constant $\hbar $, and this term is
responsible for quantum effects. The linearity of the Schr\"{o}dinger
equation (\ref{c2.2}) may be considered as a result of the special choice of
the arbitrary constant $b=\hbar $.

Such a choice is justified, because it transforms the natural dynamic
equation (\ref{c3.4}) into the linear dynamic equation (\ref{c2.2}), which
is very convenient for solution and for investigation of the dynamic system $%
\mathcal{S}_{\mathrm{S}}$. However, the dynamic equation (\ref{c2.2})
remains to be an artificial, because the linear property of dynamic equation
is not an essential property. It is appears as a result of the special
choice of the integration constant, and the linearity may not be used for
the generalization of the nonrelativistic quantum theory on the relativistic
case.

The fact that the classical approximation $\mathcal{S}_{\mathrm{cl}}$ of the
Schr\"{o}dinger particle $\mathcal{S}_{\mathrm{S}}$ is a statistical
ensemble $\mathcal{E}\left[ \mathcal{S}_{\mathrm{d}}\right] $ of free
classical (deterministic) particles $\mathcal{S}_{\mathrm{d}}$ suggests the
idea, that the Schr\"{o}dinger particle $\mathcal{S}_{\mathrm{S}}$ is in
reality a statistical ensemble $\mathcal{E}\left[ \mathcal{S}_{\mathrm{st}}%
\right] $ of free stochastic particles $\mathcal{S}_{\mathrm{st}}$. This
idea is the old reasonable idea, which has been suggested by different
authors, for instance \cite{M49}. However, the mathematical realization of
this idea met \textit{difficulties, conditioned by incorrect application of
the relativity principles}.

\section{Statistical description of relativistic particles}

Any statistical description is a description of \textit{physical objects}.
As we have mentioned above, in the nonrelativistic case the physical objects
are points of the three-dimensional space. In the relativistic case the
physical objects are lengthy objects: emlons. Statistical description of
nonrelativistic particles distinguishes from that of relativistic particles
in the sense, that the state density $\rho $ at the nonrelativistic
description is defined by the relation 
\begin{equation}
dN=\rho dV  \label{a1.1}
\end{equation}%
where $dN$ is the number of particles in the infinitesimal 3-volume $dV$. In
the relativistic case the state density $j^{k}$ is described by the relation 
\begin{equation}
dN=j^{k}dS_{k}  \label{a1.2}
\end{equation}%
where $dN$ is the flux of world lines through the infinitesimal area $dS_{k}$%
. It follows from the relations (\ref{a1.1}), (\ref{a1.2}) that in the
nonrelativistic case one can introduce the concept of the probability
density of the state on the basis of the nonnegative quantity $\rho $,
whereas in the relativistic case it is impossible, because one cannot
construct the probability density on the basis of the 4-vector $j^{k}$.

Statistical description is a description of the statistical ensemble, which
is the dynamic system consisting of many identical independent systems.
These systems may be dynamic or stochastic. However, the statistical
ensemble is a dynamic system in any case. It means, that there are dynamic
equations, which describe the evolution of the statistical ensemble state.
Investigation of the statistical ensemble as a dynamic system admits one to
investigate the mean characteristics of the stochastic systems, constituting
this ensemble. Besides, in the nonrelativistic case the statistical ensemble
is a tool for calculation of different mean quantities and distributions,
because in this case the ensemble state may be interpreted as the
probability density of the fact, that the system state is placed at some
given point of the phase space.

The statistical ensemble is used usually in the statistical physics, where
the statistical description of the deterministic nonrelativistic systems is
produced. The principal property of the statistical ensemble (\textit{to be
a dynamic system}) is perceived as some triviality, and the statistical
ensemble is considered usually as a tool for calculation of mean values of
different functions of the state. When one tries to apply this conception of
the statistical ensemble to description of relativistic stochastic
particles, it is quite natural that one fails, because the probabilistic
conception of the statistical ensemble (statistical ensemble as a tool for
calculation of mean values) cannot be applied here. The problem of
construction of the dynamic system (statistical ensemble) from stochastic
systems is not stated in the statistical physics.

We display in the example of free nonrelativistic particles, how the
statistical ensemble is constructed without a reference to the probability
theory. The action $\mathcal{A}_{\mathcal{S}_{\mathrm{d}}}$ for the free
deterministic particle $\mathcal{S}_{\mathrm{d}}$ has the form 
\begin{equation}
\mathcal{A}_{\mathcal{S}_{\mathrm{d}}}\left[ \mathbf{x}\right] =\int \frac{m%
}{2}\left( \frac{d\mathbf{x}}{dt}\right) ^{2}dt  \label{c3.7a}
\end{equation}%
where $\mathbf{x}=\mathbf{x}\left( t\right) $.

For the pure statistical ensemble $\mathcal{A}_{\mathcal{E}\left[ \mathcal{S}%
_{\mathrm{d}}\right] }$ of free deterministic particles we obtain the action 
\begin{equation}
\mathcal{A}_{\mathcal{E}\left[ \mathcal{S}_{\mathrm{d}}\right] }\left[ 
\mathbf{x}\right] =\int \frac{m}{2}\left( \frac{d\mathbf{x}}{dt}\right)
^{2}dtd\mathbf{\xi }  \label{c3.7}
\end{equation}%
where $\mathbf{x}=\mathbf{x}\left( t,\mathbf{\xi }\right) $ is a 3-vector
function of independent variables $t,\mathbf{\xi =}\left\{ \xi _{1,}\xi
_{2},\xi _{3}\right\} $. The variables (Lagrangian coordinates) $\mathbf{\xi 
}$ label particles $\mathcal{S}_{\mathrm{d}}$ of the statistical ensemble $%
\mathcal{E}\left[ \mathcal{S}_{\mathrm{d}}\right] $. The statistical
ensemble $\mathcal{E}\left[ \mathcal{S}_{\mathrm{d}}\right] $ is a dynamic
system of hydrodynamic type.

The statistical ensemble $\mathcal{E}\left[ \mathcal{S}_{\mathrm{st}}\right] 
$ of free \textit{stochastic} particles $\mathcal{S}_{\mathrm{st}}$ is a
dynamic system, described by the action 
\begin{equation}
\mathcal{A}_{\mathcal{E}\left[ \mathcal{S}_{\mathrm{st}}\right] }\left[ 
\mathbf{x,u}_{\mathrm{df}}\right] =\int \left\{ \frac{m}{2}\left( \frac{d%
\mathbf{x}}{dt}\right) ^{2}+\frac{m}{2}\mathbf{u}_{\mathrm{df}}^{2}-\frac{%
\hbar }{2}\mathbf{\nabla u}_{\mathrm{df}}\right\} dtd\mathbf{\xi }
\label{c4.1}
\end{equation}%
where $\mathbf{u}_{\mathrm{df}}=\mathbf{u}_{\mathrm{df}}\left( t,\mathbf{x}%
\right) $ is a diffusion velocity, describing the mean value of the
stochastic component of velocity, whereas $\frac{d\mathbf{x}}{dt}\left( t,%
\mathbf{\xi }\right) $ describes the regular component of the particle
velocity, and $\mathbf{x}=\mathbf{x}\left( t,\mathbf{\xi }\right) $ is the
3-vector function of independent variables $t,\mathbf{\xi =}\left\{ \xi
_{1,}\xi _{2},\xi _{3}\right\} $. The variables $\mathbf{\xi }$ label
particles $\mathcal{S}_{\mathrm{st}}$, substituting the statistical
ensemble. The operator 
\[
\mathbf{\nabla =}\left\{ \frac{\partial }{\partial x^{1}},\frac{\partial }{%
\partial x^{2}},\frac{\partial }{\partial x^{2}}\right\} 
\]%
is defined in the coordinate space of $\mathbf{x}$. Note that the transition
from the statistical ensemble (\ref{c3.7}) to the statistical ensemble (\ref%
{c4.1}) is purely dynamic. The concept of probability is not used. The
character of stochasticity is determined by the form of two last terms in
the action (\ref{c4.1}). For instance, if we replace $\mathbf{\nabla v}_{%
\mathrm{df}}$ in (\ref{c4.1}) by some function $f\left( \mathbf{\nabla v}_{%
\mathrm{df}}\right) $, we obtain another type of stochasticity, which does
not coincide with the quantum stochasticity.

The action for the single stochastic particle is obtained from the action (%
\ref{c4.1}) by omitting integration over $\mathbf{\xi }$. We obtain the
action 
\begin{equation}
\mathcal{A}_{\mathcal{S}_{\mathrm{st}}}\left[ \mathbf{x,u}_{\mathrm{df}}%
\right] =\int \left\{ \frac{m}{2}\left( \frac{d\mathbf{x}}{dt}\right) ^{2}+%
\frac{m}{2}\mathbf{u}_{\mathrm{df}}^{2}-\frac{\hbar }{2}\mathbf{\nabla u}_{%
\mathrm{df}}\right\} dt  \label{c4.1a}
\end{equation}%
were $\mathbf{x}=\mathbf{x}\left( t\right) ,\ \ \mathbf{u}_{\mathrm{df}}=%
\mathbf{u}_{\mathrm{df}}\left( t,\mathbf{x}\right) $. However, the action (%
\ref{c4.1a}) has only a symbolic sense, as far as the operator $\mathbf{%
\nabla }$ is defined in some vicinity of the point $\mathbf{x}$, but not at
the point $\mathbf{x}$ itself. It means, that the action (\ref{c4.1a}) does
not determine dynamic equations for the particle $\mathcal{S}_{\mathrm{st}}$%
, and the particle appears to be stochastic, although dynamic equations
exist for the statistical ensemble of such particles. They are determined by
the action (\ref{c4.1}). Thus, the particles described by the action (\ref%
{c4.1}) are stochastic, because there are no dynamic equations for a single
particle. In the case, when the quantum constant $\hbar =0$, the actions (%
\ref{c4.1a}) and (\ref{c3.7a}) coincide, because in this case it follows
from (\ref{c4.1a}), that $\mathbf{u}_{\mathrm{df}}=0$.

Variation of action (\ref{c4.1}) with respect to variable $\mathbf{u}_{%
\mathrm{df}}$ leads to the equation 
\begin{equation}
\mathbf{u}_{\mathrm{df}}=-\frac{\hbar }{2m}\mathbf{\nabla }\ln \rho ,
\label{c4.5}
\end{equation}
where the particle density $\rho$ is defined by the relation 
\begin{equation}
\rho =\left[ \frac{\partial \left( x^{1},x^{2},x^{3}\right) }{\partial
\left( \xi _{1},\xi _{2},\xi _{3}\right) }\right] ^{-1}=\frac{\partial
\left( \xi _{1},\xi _{2},\xi _{3}\right) }{\partial \left(
x^{1},x^{2},x^{3}\right) }  \label{c4.4}
\end{equation}
The relation (\ref{c4.5}) is the expression for the mean diffusion velocity
in the Brownian motion theory.

Eliminating $\mathbf{u}_{\mathrm{df}}$ from the dynamic equation for $%
\mathbf{x}$, we obtain dynamic equations of the hydrodynamic type. 
\begin{equation}
m\frac{d^{2}\mathbf{x}}{dt^{2}}=-\mathbf{\nabla }U\left( \rho ,\mathbf{%
\nabla }\rho \right) ,\qquad U\left( \rho ,\mathbf{\nabla }\rho \right) =%
\frac{\hbar ^{2}}{8m}\left( \frac{\left( \mathbf{\nabla }\rho \right) ^{2}}{%
\rho ^{2}}-2\frac{\mathbf{\nabla }^{2}\rho }{\rho }\right)  \label{c4.7}
\end{equation}
By means of the proper change of variables these equations can be reduced to
the Schr\"odinger equation \cite{R99}.

However, there is a serious mathematical problem here. The fact is that the
hydrodynamic equations are to be integrated, in order they can be described
in terms of the wave function. The fact, that the Schr\"{o}dinger equation
can be written in the hydrodynamic form, is well known \cite{M26,B52}.
However, the inverse transition from the hydrodynamic equations to the wave
function was not known until the end of the 20th century \cite{R99}, and the
necessity of integration of hydrodynamic equations was a reason of this fact.

Derivation of the Schr\"{o}dinger equation as a partial case of dynamic
equations, describing the statistical ensemble of random particles (\ref%
{c4.1}), shows that the wave function is simply a method of description of
hydrodynamic equations, but not a specific quantum object, whose properties
are determined by the quantum principles. At such an interpretation of the
wave function the quantum principles appear to be superfluous, because they
are necessary only for explanation, what is the wave function and how it is
connected with the particle properties. All remaining information is
contained in the dynamic equations. It appears that the quantum particle is
kind of stochastic particle, and all its exhibitions can be interpreted
easily in terms of the statistical ensemble of stochastic particles (\ref%
{c4.1}).

The idea of that, the quantum particle is simply a stochastic particle, is
quite natural. It was known many years ago \cite{M49}. However, the
mathematical form of this idea could not be realized for a long time because
of the two problems considered above (incorrect conception on the
statistical ensemble of relativistic particles and necessity of integration
of the hydrodynamic equations).

Thus, we see in the example of the Schr\"{o}dinger particle, that quantum
systems are a special sort of dynamic systems, which could be obtained from
the statistical ensemble of classical dynamic systems by means of a change
of parameters $P$ of the dynamic system by its effective value $P_{\mathrm{%
eff}}$. In particular, the free uncharged particle is described by an unique
parameter: its mass $m$.

Statistical ensemble of free classical relativistic particles is described
by the action 
\begin{equation}
\mathcal{A}_{\mathcal{E}\left[ \mathcal{S}_{\mathrm{d}}\right] }\left[ x%
\right] =-\int mc\sqrt{g_{ik}\dot{x}^{i}\dot{x}^{k}}d\tau d\mathbf{\xi
,\qquad }\dot{x}^{k}\equiv \frac{dx^{k}}{d\tau }  \label{c5.1}
\end{equation}%
where $x^{k}=x^{k}\left( \tau ,\mathbf{\xi }\right) $. To obtain the quantum
description, we are to consider the statistical ensemble $\mathcal{E}\left[ 
\mathcal{S}_{\mathrm{st}}\right] $ of free stochastic relativistic particles 
$\mathcal{S}_{\mathrm{st}}$, which is the dynamic system described by the
action%
\begin{equation}
\mathcal{A}_{\mathcal{E}\left[ \mathcal{S}_{\mathrm{st}}\right] }\left[ x,u%
\right] =-\int m_{\mathrm{eff}}c\sqrt{g_{ik}\dot{x}^{i}\dot{x}^{k}}d\tau d%
\mathbf{\xi ,\qquad }\dot{x}^{k}\equiv \frac{dx^{k}}{d\tau }  \label{c5.2}
\end{equation}%
where $x^{k}=x^{k}\left( \tau ,\mathbf{\xi }\right) $, $u^{k}=u^{k}\left(
x\right) $, $k=0,1,2,3$. Here the effective mass is obtained from the mass $m
$ of the deterministic (classical) particle by means of the change%
\begin{equation}
m^{2}\rightarrow m_{\mathrm{eff}}^{2}=m^{2}\left( 1+g_{ik}\frac{u^{i}u^{k}}{%
c^{2}}+\frac{\hbar }{mc^{2}}\partial _{k}u^{k}\right)   \label{c5.3}
\end{equation}%
where $u^{k}=u^{k}\left( x\right) $ the mean value of the 4-velocity
stochastic component. Using the relation%
\begin{equation}
\kappa ^{k}=\frac{m}{\hbar }u^{k},  \label{c5.4}
\end{equation}%
it is convenient to introduce the 4-velocity $\kappa =\left\{ \kappa ^{0},%
\mathbf{\kappa }\right\} $, with $\mathbf{\kappa }$ having dimensionality of
the length. The action (\ref{c5.2}) takes the form%
\begin{equation}
\mathcal{A}_{\mathcal{E}\left[ \mathcal{S}_{\mathrm{st}}\right] }\left[
x,\kappa \right] =-\int mcK\sqrt{g_{ik}\dot{x}^{i}\dot{x}^{k}}d\tau d\mathbf{%
\xi ,\qquad }K\mathbf{=}\sqrt{1+\lambda ^{2}\left( g_{ik}\kappa ^{i}\kappa
^{k}+\partial _{k}\kappa ^{k}\right) }  \label{c5.5}
\end{equation}%
where $\lambda =\frac{\hbar }{mc}$ is the Compton wave length of the
particle and the metric tensor $g_{ik}=$diag$\left\{ c^{2},-1,-1,-1\right\} $%
. In the nonrelativistic approximation the action (\ref{c5.5}) turns in the
action (\ref{c4.1}), which takes the form%
\begin{equation}
\mathcal{A}_{\mathcal{S}_{\mathrm{st}}}\left[ \mathbf{x,u}\right] =\int
\left\{ -mc^{2}+\frac{m}{2}\left( \frac{d\mathbf{x}}{dt}\right) ^{2}+\frac{m%
}{2}\mathbf{u}^{2}-\frac{\hbar }{2}\mathbf{\nabla u}\right\} dtd\mathbf{\xi }
\label{c5.6}
\end{equation}%
Deriving (\ref{c5.6}), we choose the parameter $\tau =t=x^{0}$, take into
account the relation (\ref{c5.4}) and neglect $\partial _{0}\kappa ^{0}$ as
compared with $\mathbf{\nabla \kappa }$.

In the general case (\ref{c5.5}) the $\kappa $-field $\kappa ^{k}$ may be
also represented in the form of gradient as well as in the case (\ref{c4.5})%
\begin{equation}
\kappa ^{k}=g^{kl}\partial _{l}\kappa  \label{c5.7}
\end{equation}%
Here $\kappa $ is the scalar field of such a form, that $e^{\kappa }$
satisfies the inhomogeneous wave equation.

Using proper change of variables, one can introduce the wave function,
satisfying the Klein-Gordon equation. At such a change of variables the $%
\kappa $-field appears to be hidden in the wave function and its remarkable
properties appear to be hidden. As well as the diffusion velocity $\mathbf{u}%
_{\mathrm{df}}$ in (\ref{c4.1a}) the $\kappa $-field $\kappa ^{k}$ describes
the mean value of the stochastic component of the particle velocity. In the
non-relativistic case (\ref{c4.1a}) the 3-velocity $\mathbf{u}_{\mathrm{df}}$
is determined uniquely by its source (the density $\rho $ of particles). But
the $\kappa $-field is a relativistic field, which may escape from its
source and exist separately from its source. Besides, the $\kappa $-field
can change the effective particle mass, as one can see from the expression (%
\ref{c5.5}) for the action. In particular, if 
\begin{equation}
\lambda ^{2}\left( g_{ik}\kappa ^{i}\kappa ^{k}+\partial _{k}\kappa
^{k}\right) <-1,\qquad K^{2}<0  \label{c5.8}
\end{equation}%
the particle mass becomes imaginary. In this case the mean world line of the
particle is spacelike, and the pair production becomes to be possible. In
other words, the $\kappa $-field can produce pairs \cite{R2003}.

The property of pair production is a crucial property of the quantum
relativistic physics. The classical fields (electromagnetic, gravitational)
do not possess this property. As we have seen in the second section the
description in framework of the conventional QFT has problems with the pair
production description. There is a hope, that the proper statistical
description of several relativistic stochastic particle will admit one to
obtain the pair production effect. For instance, maybe, two colliding
relativistic particles can produce pairs by means of their common $\kappa $%
-field. We hope that such a program may appear to be successful, provided
the proper formalism of the statistical ensemble will be developed. Today we
have only the developed formalism for statistical description of stochastic
system consisting of one emlon (WL). Formalism for statistical description
of stochastic system consisting of several emlons (WLs) is not yet developed
properly.

\section{Epistemological problems of quantum theory}

Construction of the relativistic quantum theory is a very difficult problem.
But solution of this problem depends not only on the difficulty of the
problem in itself. It depends also on qualification of researchers,
investigating this problem, on effectiveness of the applied investigation
methods, on capability of researchers to logical reasonings and on other
factors. In this section we shall try analyze the character of appearing
difficulties. We shall separate them into two parts: objective difficulties
and subjective difficulties. The objective difficulties have been discussed.
Further we shall try to discuss subjective difficulties and mistakes.
Discovery and correction of these mistakes is difficult because of their
subjective character.

In our opinion, the main difficulty is a deficit of logic (predominance of
the trial and error method over logic) at the construction of the quantum
relativistic theory. In particular, this deficit of logic is displayed in
disregard of demands, imposed by the relativity principles. Let us consider
briefly the history of the question. In the beginning of the 20th century
there were attempts of constructon of the nonrelativistic quantum mechanics
as a statistical description of stochastic particles. In these attempts the
statistical description was considered to be a probabilistic description.
Incompatibility of the probabilistic description with the relativity
principles was not known, because one ignored the circumstance that the
world line (but not the particle) was the physical object. Because of this
mistake one could not construct the statistical conception of the quantum
mechanics. One succeeded to construct the axiomatic conception of quantum
mechanics by means of the trial and error method. After this success the
trial and error method became the principal investigation method in the
quantum theory. The trial and error method had the success and became to be
predominant, because it was insensitive to mistakes in the fundamental
physical principles, whereas the classical investigation method, which run
back to Isaac Newton, was founded on the logic. The method, founded on logic
could not lead to correct results, if the fundamental physical principles
contained mistakes, or these principles were applied incorrectly.

In the first half of the 20th century there were classics of physics, who
knew and used the classical logical method of investigation. In the last
half of the 20th century, there were only researchers, using the trial and
error method. The predominancy of the trial and error method is explained by
two factors. Firstly, it appeared to be successful in application to
construction of the nonrelativistic quantum theory. Secondly, the classical
logical method appeared to be forgotten, because new generations of the
researcher were educated on the example of the successful application of the
trial and error method, which was perceived as the only possible method of
investigation. Any ambitious theorist dreamed to invent such hypothesis
(maybe, very exotic), which could be explain at once the mass spectrum of
elementary particles and solve other problems of QFT. Development of the
microcosm physics turned into competition of such hypotheses.

The circumstance, that the correct application of the relativity principles
(the correct application of the fact that the world line is a physical
object) may be important from the practical viewpoint became to be clear for
the author of this paper after investigation of the world line properties 
\cite{R70}. Two important results followed from this paper: (1) the quantum
mechanics as a statistical theory can be constructed, if one uses the
relativistic concept of the state and construct the statistical description
without a use the probability theory \cite{R71,R73b,R73a}, (2) the
perturbation theory at the second quantization may be eliminated, if the
conservation law of physical objects (world lines) is taken into account) 
\cite{R72}. The paper \cite{R70} was reported at the seminar of the
theoretical department of the Lebedev physical institute. Relation to the
paper was sceptical as far as results were obtained without any additional
suppositions (and this was unusual). Besides, many researchers did not
believe, that it was possible a classical description of the world line,
making a zigzag in the time direction. Further the calculation were tested,
and all objections were eliminated.

It was clear to the author of paper \cite{R70}, that the first and the
second results were incompatible. He believed that the quantum mechanics is
the statistical description of random world lines and the quantum principles
are to be corollaries of this description. However, in that time the
integration of hydrodynamic equations was not known, and from the
mathematical viewpoint the statistical description could not be considered
as a starting point of the quantum description. The second result admitted
one to separate the problem of the second quantization in to parts and to
solve exactly the one-emlon problem and the two-emlon problem without a use
of the perturbation theory. The author expected that the further development
of the second quantization led the problem into the blind alley, provided
the fitting is not used. From his viewpoint it should prove that development
of QFT in the direction of the second quantization were erroneous.

No additional hypotheses were used, to avoid a charge in a use of erroneous
hypotheses, leading to a strange result (absence of the pair production). In
particular, one uses neither the perturbation theory, nor cut off the
self-action at the time tending to infinity. (Usually the two hypotheses are
always used). Under these conditions the absence of the pair productions
meant, that \textit{the strategy of the second quantization in itself is
erroneous}, as far as the relativistic quantum field theory, where there is
no pair production, cannot be true. When the corresponding paper was
submitted to a scientific journal, it was rejected on the basis of the
review of the referee, who wrote: " I do not recommend the paper for
publication, because the author himself stated that in his method of
quantization the pair production is absent." (The paper has not been
published, and now it can be found only at the site \cite{R90a})

We see here a sample of logic, based on the trial and error method, which
does not accept the papers with a negative result. The referee does not
admit existence of other investigation methods other, than the trial and
error method. Indeed, as far as in the trial and error method all hypotheses
are random, the tests leading to a negative result are of no interest. In
the logical investigation method the negative result means that the starting
premises are false (of course, if there are no mistakes in the executed
investigation). Unfortunately, the approach of the referee is typical. Most
researchers are apt to use only the trial and error method and they cannot
imagine anything other than this method. During thirty years the author of
this paper had a chance to discuss the correctness of the second
quantization problem with his colleagues dealing with QFT. Some of them
agreed that, maybe, the commutation relation are incompatible with the
dynamic equations. But at the same time they stated that it means nothing,
because QFT agrees with the experimental data very well. The circumstance,
that the experimental data are explained by means of the inconsistent theory
did not lead to objections from them. Such an approach is a corollary of the
predominant method of trial and error. We think that this method is the main
obstacle on the path of the relativistic quantum theory construction. 
\textit{One can find and correct mistakes in the theory, but a change of
mentality needs some time. This time may be rather long}.

We have seen that the nonrelativistic quantum theory could be presented as a
statistical description of stochastic particles, if we apply the relativity
principles correctly and use the \textit{dynamic conception} of the
statistical description. In fact, the nonrelativistic quantum theory was
developed mainly by the trial and error method. Appearance of quantum
principles was a result of application of this method. The trial and error
method is an effective method for investigation of single physical phenomena
of unknown nature, because it admits one to discover new concepts, which are
adequate to the considered phenomenon. However, the trial and error method
is inadequate for construction of a fundamental theory, because the
fundamental theory is a logical structure, which systematizes our knowledge.
The systematization needs a long logical reasonings, because it is based on
several fundamental propositions. The systematization as well as the
fundamental theory is very sensitive to possible mistakes in the logical
reasonings and in the mathematical calculations. Any mistake should be
analyzed and eliminated.

On the contrary, the trial and error method is insensitive to mistakes. It
is multiple-path. Usually before obtaining a correct solution one suggests
and tests many different hypotheses. Only one of them may appear to be true.
As far as the hypotheses are suggested occasionally, it is useless to
analyze the erroneous hypotheses. Such an analysis gives nothing, because
the hypotheses do not connect between themselves and with the obtained
correct result.

If we use logical reasonings, based on the fundamental principles, and
obtain an incorrect result, it means that either we make a mistake, or the
fundamental principles are incorrect. Thus, at the logical approach we
should discover and analyze our mistakes. It is useful for a correction of
our reasonings. At the engineering approach to the construction of the
fundamental theory, when one uses the trial and error method, a discovery
and an analysis of the possible mistake is useless. Furthermore, it is
possible such a case, when the obtained result is incorrect, although it
agrees with the experimental data. Let us illustrate this in the example,
which relates to the problem of the relativistic quantum theory construction.

We discuss the problem, whether the Dirac particle $\mathcal{S}_{\mathrm{D}}$
is pointlike or it has an internal structure. The Dirac particle $\mathcal{S}%
_{\mathrm{D}}$ is the dynamic system, described by the action of the form%
\begin{equation}
\mathcal{S}_{\mathrm{D}}:\qquad \mathcal{A}_{\mathrm{D}}[\bar{\psi},\psi
]=c^{2}\int (-mc\bar{\psi}\psi +\frac{i}{2}\hbar \bar{\psi}\gamma
^{l}\partial _{l}\psi -\frac{i}{2}\hbar \partial _{l}\bar{\psi}\gamma
^{l}\psi -\frac{e}{c}A_{l}\bar{\psi}\gamma ^{l}\psi )d^{4}x  \label{b5.1}
\end{equation}%
where $m$ and $e$ are respectively the mass and the charge of the Dirac
particle, and $c$ is the speed of the light. Here $\psi $ is four-component
complex wave function, $\psi ^{\ast }$ is the Hermitian conjugate wave
function, and $\bar{\psi}=\psi ^{\ast }\gamma ^{0}$ is the conjugate one.
The quantities $\gamma ^{i}$, $i=0,1,2,3$ are $4\times 4$ complex constant
matrices, satisfying the relation 
\begin{equation}
\gamma ^{l}\gamma ^{k}+\gamma ^{k}\gamma ^{l}=2g^{kl}I,\qquad k,l=0,1,2,3.
\label{b5.2}
\end{equation}%
where $I$ is the $4\times 4$ identity matrix, and $g^{kl}=$diag$\left(
c^{-2},-1,-1,-1\right) $ is the metric tensor. The quantity $A_{k}$, $%
k=0,1,2,3$ is the electromagnetic potential. The action (\ref{b5.1})
generates the dynamic equation for the dynamic system $\mathcal{S}_{\mathrm{D%
}}$, known as the Dirac equation 
\begin{equation}
\gamma ^{l}\left( -i\hbar \partial _{l}+\frac{e}{c}A_{l}\right) \psi +mc\psi
=0  \label{b5.3}
\end{equation}%
and expressions for physical quantities: the 4-flux $j^{k}$ of particles and
the energy-momentum tensor $T_{l}^{k}$%
\begin{equation}
j^{k}=c^{2}\bar{\psi}\gamma ^{k}\psi ,\qquad T_{l}^{k}=\frac{ic^{2}}{2}%
\left( \bar{\psi}\gamma ^{k}\partial _{l}\psi -\partial _{l}\bar{\psi}\cdot
\gamma ^{k}\psi \right)   \label{b5.4}
\end{equation}

If the Dirac particle $\mathcal{S}_{\mathrm{D}}$ is not pointlike and has an
internal structure, described by some additional degrees of freedom, this
structure is to be present also in the nonrelativistic approximation.
Conventionally one assumes that the Pauli particle $\mathcal{S}_{\mathrm{P}}$
is the nonrelativistic approximation of the Dirac particle $\mathcal{S}_{%
\mathrm{D}}$ (see, for instance, \cite{D58}).

In the partial case, when $A_{0}=0$, the Pauli particle $\mathcal{S}_{%
\mathrm{P}}$ is the dynamic system, described by the dynamic equation 
\begin{equation}
i\hbar \partial _{0}\psi _{1}=\left( \frac{\pi _{\mu }\pi _{\mu }}{2m}+\frac{%
ie\hbar }{2mc}\varepsilon _{\nu \mu \alpha }\partial _{\nu }A_{\mu }\sigma
_{\alpha }\right) \psi _{1},\qquad \pi _{\alpha }\equiv i\hbar \partial
_{\alpha }-\frac{e}{c}A_{\alpha },\qquad \alpha =1,2,3  \label{b5.5}
\end{equation}%
where $\psi _{1}$ is the two-component complex wave function, $\varepsilon
_{\mu \nu \alpha }$ is the Levi-Chivita pseudotensor, and $\mathbf{\sigma }%
=\left\{ \sigma _{1},\sigma _{2},\sigma _{3}\right\} $ are the $2\times 2$
Pauli matrices.

The Pauli particle $\mathcal{S}_{\mathrm{P}}$ is the pointlke particle,
which has no internal structure. This fact agrees with the experimental
data. Hence, if the Pauli particle $\mathcal{S}_{\mathrm{P}}$ is the
nonrelativistic approximation of the Dirac particle $\mathcal{S}_{\mathrm{D}%
} $, the Dirac particle is pointlike also and has no internal structure. The
Pauli equation (\ref{b5.5}) has a lower order (four first order real
equations), than the Dirac equation (\ref{b5.3}) (eight first order real
equations). It means that at transition from the Dirac equation to the Pauli
equation the order of the system of dynamic equation reduces, and several
degrees of freedom were lost.

The equation (\ref{b5.5}) is obtained from equation (\ref{b5.3}) as the
limit $c\rightarrow \infty $. Some temporal derivatives $\partial _{0}$ in
the Dirac equation (\ref{b5.3}) have small coefficients of the order $c^{-1}$
and $c^{-2}$. These terms are neglected and the order of the system of
dynamic equations reduces. However, the neglected terms are the terms with
the small parameters before the highest derivative. One cannot neglect these
terms, because at very high frequencies (of the order $\Omega =mc^{2}/\hbar $%
) these terms become to be of the same order as the remaining terms.
Neglecting these terms, we neglect the high frequency degrees of freedom.

In reality, the states of the Dirac particle, which are linear superposition
of conventional low frequency state with the high frequency state are
unstable, because in this case the 4-current $j^{k}=c^{2}\bar{\psi}\gamma
^{k}\psi $ oscillates with the frequency of the order $\Omega =mc^{2}/\hbar $%
. The Dirac particle is charged. As a result the energy of the high
frequency excitation is radiated in the form of electromagnetic waves, and
the Dirac particle appears at the state, where the 4-current $j^{k}=c^{2}%
\bar{\psi}\gamma ^{k}\psi $ is constant. Thus, from the experimental
viewpoint the additional high frequency degrees of freedom of the Dirac
particle do not exist, because they are not observable.

Can one discover these degrees of freedom theoretically from the analysis of
the Dirac particle? Yes, one can discover them at the scrupulous analysis in
the framework of the conventional quantum mechanics \cite{R2005a}. But they
have not been discovered. We are not sure, whether the theory of
differential equations with small parameter before the highest derivative
was known in the first half of the 20th century, but it was definitely known
in the last half of the 20th century. Nevertheless the necessary analyses
has not been produced, and the Dirac particle was considered to be pointlike.

It is true that the high frequency degrees of freedom are not observable at
low energies, and they give no contribution to description of the Dirac
particle in the nonrelativistic case. One can assume that these degrees of
freedom absent in the nonrelativistic case, and this assumption agrees with
the experimental data. However, at the high energy collisions of Dirac
particles these degrees of freedom may be excited and make an essential
contribution to description of the high energy collision process.

Why has one not discovered these degrees of freedom theoretically? Because
researchers used the trial and error method, where the only criterion of
validity of the theory is the agreement with experiment. The logical
reasonings and mistakes in consideration are of no importance, provided they
do not violate agreement with experiment. If we take into account that the
nonrelativistic quantum theory was created by means of the trial and error
method, and three generations of the microcosm researchers have been
educated on the example of this method, we recognize that the internal
degrees of freedom of the Dirac particle could not be discovered before the
execution of the proper high energy experiments with Dirac particles.

The internal degrees of freedom of the Dirac particle were discovered at
investigation of the Dirac particle by dynamic methods \cite{R2004a}, which
use the logical approach to investigation. The dynamic methods are attentive
to the logic and to mistakes of investigation. They are not oriented to the
trial and error method and to agreement with experiment.

Besides, investigation of the Dirac equation by the dynamic methods has
shown \cite{R2004b} that the description of internal degrees of freedom is
nonrelativistic. It means that the whole Dirac equation (\ref{b5.1}) is a
nonrelativistic equation, although it is written in the relativistically
covariant form. Nonrelativistic character of the Dirac equation means
mathematically, that the set of all solutions of the Dirac equation is not
invariant, in general, under some Poincar\'{e} transformations.

From viewpoint of researchers, who believed only in experimental test (but
not in logic reasonings) the Dirac equation is relativistic, because only
internal degrees of freedom are described nonrelativistically, but these
degrees of freedom are ignored at the conventional approach. Publishing of
the papers \cite{R2004a,R2004b} in the reviewed journal appears to be
impossible, because the referee declared that he cannot believe that the
Dirac equation be nonrelativistic. This declaration appears to be sufficient
for rejection of the paper. I think that the opinion of the referee reflects
the viewpoint of the statistical average researcher, and this opinion is
erroneous.

Problem of the relativistic invariance of the Dirac equation is discussed in 
\cite{R2004b} in details. Here we shall not go in details. We refer only to
the theorem, formulated by Anderson \cite{A67}. This theorem states: \textit{%
The symmetry group of dynamic equations, written in the relativistically
covariant form, coincides with the symmetry group of absolute objects.} The
absolute objects are quantities, which are the same for all solutions of the
dynamic equations. In the Dirac equation (\ref{b5.3}) the quantities $A_{k}$
and matrices $\gamma ^{k}$ are absolute objects. We set for simplicity $%
A_{k}=0$. Then the symmetry group of 4-vector $A_{k}=0$ and of the unit
matrix 4-vector $\gamma ^{k}$ is the group of translation and the group of
rotation around the direction defined by the unit 4-vector $\gamma ^{k}$.
This 7-parametric group is a subgroup of the 10-parametric Poincar\`{e}
group. Hence, the Dirac equation is not relativistic.

The Dirac equation distinguishes from the Klein-Gordon equation in the
sense, that the Klein-Gordon equation contains the absolute object $g^{kl}=$%
diag$\{c^{-2},-1,$ $-1,-1\}$, which has the 10-parametric Poincar\`{e} group
as its symmetry group.

As an illustration of the role of unit constant vector in the
relativistically covariant equation, we note the dynamic equation for the
free nonrelativistic classical particle 
\begin{equation}
m\frac{d^{2}\mathbf{x}}{dt^{2}}=0  \label{b5.6}
\end{equation}%
can be written in the relativistically covariant form, if one introduces the
unit timelike constant 4-vector $l_{k}$ $\ \ (l_{k}g^{kl}l_{l}=1)$. We
obtain instead of (\ref{b5.6})%
\begin{equation}
\frac{mc^{2}}{\left( l_{n}\dot{x}^{n}\right) }\frac{d}{d\tau }\left[ \frac{%
\dot{x}^{i}}{\left( l_{k}\dot{x}^{k}\right) }-{\frac{1}{2}}g^{ik}l_{k}\frac{%
\dot{x}^{s}g_{sl}\dot{x}^{l}}{\left( l_{j}\dot{x}^{j}\right) ^{2}}\right]
=0,\qquad \dot{x}^{i}\equiv \frac{dx^{i}}{d\tau },\qquad i=0,1,2,3
\label{b5.7}
\end{equation}%
Indeed, setting $l_{k}=\left\{ c,0,0,0\right\} $ in (\ref{b5.7}), we obtain
for $i=1,2,3$ the equation (\ref{b5.6}), because $\left( l_{k}\dot{x}%
^{k}\right) d\tau =cdt=cdx^{0}$. For $i=0$ we obtain the identity 
\[
mc^{2}\frac{d}{cdt}\left[ \frac{dt}{cdt}-{\frac{1}{2c}}\right] \equiv 0 
\]

The Newtonian space of events contains two invariants $dt=dx^{0}$ and $dr=%
\sqrt{d\mathbf{x}^{2}}$, whereas the Minkowski space-time contains only one
invariant $ds=\sqrt{c^{2}dt^{2}-dr^{2}}$. Introduction of the constant unit
4-vector $l_{k}$ admits one to construct two invariants $dt$ and $dr$ from
one invariant $ds$ and 4-vector $dx^{k}$ by means of relations 
\[
cdt=l_{k}dx^{k},\qquad dr=\sqrt{c^{2}dt^{2}-ds^{2}}=\sqrt{\left(
l_{k}dx^{k}\right) ^{2}-g_{ik}dx^{i}dx^{k}} 
\]

Thus, if the dynamic equations written in the relativistically covariant
form contain the unit timelike constant vector, we should suspect that
dynamic equations are not relativistic.

\section{Necessity of the next modification of the space-time model}

Thus, the quantum mechanics can be founded as a mechanics of stochastic
particles. However, it is not known, why the motion of free particles is
stochastic and from where the quantum constant does appear. There are two
variants of answer to these questions.

1. The stochasticity of the free particle motion is explained by the
space-time properties, and the quantum constant is a parameter, describing
the space-time properties.

2. The stochasticity of the free particle motion is explained by the special
quantum nature of particles. The motion of such a particle distinguishes
from the motion of usual classical particle. There is a series of rules
(quantum principles), determining the quantum particle motion. The universal
character of the quantum constant is explained by the universality of the
quantum nature of all particles and other physical objects. As to event
space, it remains to be the same as at Isaac Newton.

It is quite clear that the first version of explanation is simpler and more
logical, as far as it supposes \textit{only a change of the space-time
geometry}. The rest, including the principles of classical physics, remains
to be unchanged. The main problem of the first version was an absence of the
space-time geometry with such properties. In general, one could not imagine
that such a space-time geometry can exist. As a result in the beginning of
the 20th century one choose the second version. After a large work the
necessary set of additional hypotheses (quantum principles) had been
invented. One succeeded to explain all nonrelativistic quantum phenomena.
However, an attempt of the quantum theory expansion to the relativistic
phenomena lead to the problem, which is formulated as \textit{join of
nonrelativistic quantum principles with the principles of the relativity
theory}.

In general, the question, why the motion of microparticles is stochastic,
does not relate directly to the problem of the relativistic quantum theory
construction. It relates only in the sense, that explanation of the
stochasticity by the space-time properties creates an entire picture of the
world, where the good old classical principles rule, and only the space-time
properties are slightly changed. It is clear, that explanation of quantum
properties by a slight correction of the space-time properties is more
attractive, than the substitution of principles of classical physics by
enigmatic quantum principles, which are incompatible with the relativity
principles.

Besides, the correction of the space-time properties is very simple. It does
not demand an introduction of additional exotic space-time properties such
as a space-time stochasticity, or noncommutativity of coordinates in the
space-time.

Correction of the space-time properties means a change of the world function 
$\sigma $ \cite{S60} of the space-time. It consists of three points \cite%
{R90,R2002,R2005d}.

\begin{enumerate}
\item One proves that the proper Euclidean geometry has the $\sigma $%
-immanence property. It means that the proper Euclidean geometry is
described entirely by its world function $\sigma _{\mathrm{E}}$, and all
Euclidean prescriptions for construction of geometrical objects and
relations between them can be expressed in terms and only in terms of the
Euclidean world function $\sigma _{\mathrm{E}}$.

\item It is supposed that any space-time geometry $\mathcal{G}$ has the $%
\sigma $-immanence property. It means that all prescriptions of the geometry 
$\mathcal{G}$ for construction of geometrical objects and relations between
them can be obtained from the Euclidean prescription by a proper deformation
of the Euclidean geometry, i.e. by the change of the Euclidean world
function $\sigma _{\mathrm{E}}$ by the world function $\sigma $ of the
space-time geometry $\mathcal{G}$ in all Euclidean prescriptions.

\item The world function $\sigma _{\mathrm{d}}$ of the space-time geometry $%
\mathcal{G}_{\mathrm{d}}$ is chosen in the form 
\begin{equation}
\sigma _{\mathrm{d}}=\sigma _{\mathrm{M}}+D\left( \sigma _{\mathrm{M}%
}\right) ,\qquad D\left( \sigma _{\mathrm{M}}\right) =\left\{ 
\begin{array}{c}
\frac{\hbar }{2bc},\quad \text{if}\quad \sigma _{\mathrm{M}}>\frac{\hbar }{%
2bc} \\ 
0,\quad \text{if}\quad \sigma _{\mathrm{M}}<0%
\end{array}%
\right.  \label{b6.1}
\end{equation}%
where $\sigma _{\mathrm{M}}$ is the world function of the Minkowski space, $%
c $ is the speed of the light and $b\leq 10^{-17}$g/cm is the constant,
describing connection between the geometric mass $\mu $ and usual mass $m$
by means of the relation $m=b\mu $.
\end{enumerate}

In the space-time with nonvanishing distortion $D\left( \sigma _{\mathrm{M}%
}\right) $ the particle mass is geometrized \cite{R91}, and motion of free
particles is stochastic. The distortion function $D\left( \sigma _{\mathrm{M}%
}\right) $ describes the character of quantum stochasticity. Form of the
distortion function $D\left( \sigma _{\mathrm{M}}\right) $ is determined by
the demand that the stochasticity generated by distortion is the quantum
stochasticity, i.e. the statistical description of the free stochastic
particle motion is equivalent to the quantum description in terms of the Schr%
\"{o}dinger equation \cite{R91}.

\section{Concluding remarks}

We have considered two possible strategies of the relativistic quantum
theory construction. The first strategy, founded on the application of the
conventional quantum technique to relativistic systems, leads either to
inconsistent conception or to the consistent theory, where the pair
production does not appear for interactions of the degree type.

The second strategy is founded on the construction of the fundamental
theory, which relates to the conventional nonrelativistic quantum theory
approximately in such a way as the statistical physics relates to the
axiomatic thermodynamics. The fundamental theory is the conventional
relativistic classical theory in the space-time, whose geometry is slightly
modified in such a way, that motion of free particles is primordially
stochastic and the particle mass is geometrized. The quantum constant
appears as a parameter of the space-time geometry. Statistical description
of stochastic nonrelativistic particle motion appears to be equivalent to
the conventional quantum description. There is no necessity to postulate the
quantum principles, because they may be obtained as a corollary of such a
statistical description of nonrelativistic stochastic particles.

There is a hope that the direct application of the statistical description
to relativistic stochastic systems admits one to construct the relativistic
quantum theory. The fundamental theory admits one to use only the logical
investigation method of Isaac Newton. The fundamental theory is free of
application of the trial and error method, which is the main obstacle on the
path of the relativistic quantum theory construction. Predominance of the
trial and error method in the 20th century generated a specific mentality of
contemporary researchers, when the researcher tries to suggest new
hypotheses and to guess the result but not to derive it by the logical way
from the fundamental physical principles. This mentality is a very serious
obstacle on the path of the relativistic quantum theory construction.

\end{document}